\documentclass[preprint,12pt]{elsarticle}

\usepackage{graphicx}
\usepackage{amssymb}

\usepackage{lineno}

\biboptions{sort&compress}

\journal{Nuclear Instruments and Methods, Section A}

\begin{document}

\begin{frontmatter}


\title{Evidence for polyimide redeposition and possible correlation with sparks in Gas Electron Multipliers working in CF$_4$ mixtures}



\author[usp]{Thiago B. Saramela}
\author[usp]{Tiago F. Silva\fnref{myfootnote1}}
\fntext[myfootnote1]{Corresponding author. e-mail: tfsilva@usp.br}
\author[usp]{Marco Bregant}
\author[usp]{\\Marcelo G. Munhoz}
\author[uon1]{Tien T. Quach}
\author[uon1]{Richard Hague} 
\author[npl]{Ian S. Gilmore}
\author[uon2]{Clive J. Roberts} 
\author[npl,uon1]{Gustavo F. Trindade\fnref{myfootnote2}}
\fntext[myfootnote2]{Corresponding author. e-mail: gustavo.trindade@npl.co.uk}
\address[usp]{Instituto de Física da Universidade de S\~ao Paulo, Rua do matão, 1371, 05508-090 S\~ao Paulo, Brazil.}
\address[npl]{National Physical Laboratory, Hampton Road, TW11 0LW, United Kingdom }
\address[uon1]{University of Nottingham, Nottingham, NG7 2RD, United Kingdom  }
\address[uon2]{School of Life Sciences, University of Nottingham, Nottingham, NG7 2RD, United Kingdom  }

\begin{abstract}

Research on aging processes of Gas Electron Multipliers (GEMs) is important to obtain insights on how to increase the detector's longevity, stability, and performance, as highlighted in the latest developments roadmap by the European Council of Future Accelerators (ECFA). One key aspect of the aging process is the deposit formation on the electrodes surfaces. In this work, through the analysis of the molecular content on the surface of a used GEM, we provide evidence for polyimide redeposition as a source of organic material contributing to the formation of insulating layers on the electrodes, which eventually lead to sparks and detector failure. Furthermore, we show that chromium, used to promote adhesion between copper and polyimide, in the device undergoes a diffusion process, effectively blurring the layered structure. We demonstrate the significance of surface-sensitive chemical analysis to investigate the surface deposits on electrodes of gaseous detectors and our results reveal the necessity of standardization and more stringent study protocols. 

\end{abstract}

\begin{keyword}
Gas Electron Multiplier \sep Aging of gaseous detectors \sep Radiation effects \sep Surface chemical analysis \sep Secondary ion mass spectrometry


\end{keyword}

\end{frontmatter}


\section{Introduction}
\label{intro}

    The Gas Electron Multiplier (GEM) is a new technology of gaseous detectors \cite{sauli_gem:_1997, sauli_micro-pattern_2002, sauli_gas_2016} that is replacing the previous Multi-Wire Proportional Chambers (MWPC) with advantages of higher count-rates capabilities, reduced ion backflow, superior radiation hardness, among others \cite{bressan_high_1999, bachmann_performance_2001, sauli_electron_2003}. GEMs have been widely adopted in on going experiments in particle and high-energy physics, and are part of conceptual design of future experiments \cite{altunbas_construction_2002, and_upgrade_2014, collaboration_upgrade_2021, abbaneo_study_2014, ogawa_performance_2017, schueler, bortone, driuk, ciraldo, murakami}.

\begin{figure}[htb!]
        \centering
        \includegraphics[width=0.45\textwidth]{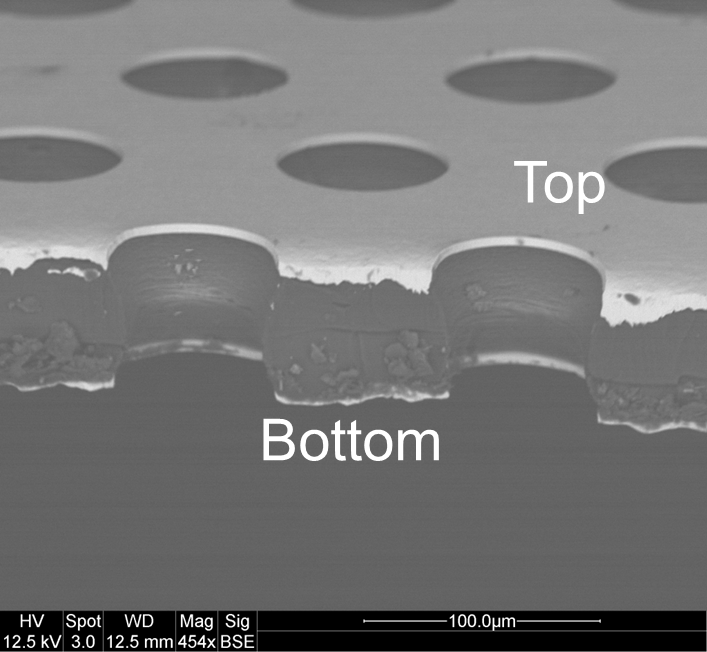}
        \caption{Scanning electron microscopy of a sectioned GEM foil showing the insulating layer with copper clad on the top and at the bottom. }
        \label{fig:foil}
    \end{figure}

    A GEM consists of a thin insulating foil (usually made of polyimide) sandwiched between two electrodes (usually made of copper) and with holes etched through it (see Fig. \ref{fig:foil}). When a voltage is applied across the two electrodes, a high electric field is generated within the holes in the insulating foil. The high electric field amplifies the ionization charge produced by radiation into a gaseous media by accelerating  the electrons, causing them to collide with other gas molecules present in the holes, and creating a cascade effect. This cascade results in the production of additional electrons, amplifying the original signal that becomes measurable. 
    GEM based detectors can be constructed with multiple layers of GEM foils stacked together, each providing additional amplification. This allows for even greater sensitivity in detecting radiation \cite{sauli_gem:_1997, sauli_micro-pattern_2002, sauli_gas_2016}.
    
    Even though the GEM radiation hardness has been shown as superior \cite{altunbas_aging_2003}, long-term aging studies of GEMs are highly important. They were considered as one of the critical points of development in the latest ECFA report \cite{ecfa_detectors_rd_roadmap_process_group_2021_2021}, affecting the longevity of many experiments. Some correlation of aging and high count rates has been reported \cite{azmoun_study_2006}.
    Thus, both, longevity and count rate capability, are constraints that can be an obstacle to reaching the physical goals of experiments.
    
    Various factors can affect or initiate the aging process of a GEM. The choice of materials for the construction of detectors, such as casings, seals, and glues, among others, can have an important influence \cite{bouclier_effects_1994, niebuhr_aging_2006, azmoun_study_2006}. It is also known, for example, that the choice of gas pipe material irreversibly affects the performance of the gaseous detector \cite{vavra_review_1986}. Even water from the environment diffused into the constituent materials has been subject of study \cite{benussi_characterization_2018} since it was reported that there is some acceleration in the aging process with humidity and correlation with sparks on GEMs \cite{azmoun_study_2006, sauli_fundamental_2003}. Separately, the erosion of the internal wall of the insulating layer in the GEM holes was reported in gas mixtures containing argon, carbon dioxide, and carbon tetrafluoride \cite{alfonsi_studies_2005}. Therefore, a deep understanding of  aging of GEMs can provide insights on choice of materials and how to meet the instruments constraints to achieve the physical results. 

    Aging of gaseous detectors has been subject of study for a long time \cite{hohlmann_aging_2002, sauli_fundamental_2003}. It can be linked to the effects of chemical reactions that occur in regions close to the electrodes that lead to the formation of insulating deposits on the conductive surface \cite{hohlmann_aging_2002, vavra_physics_2003}. Such deposits can affect gain, resolution \cite{deptuch_selected_2006}, and introduce effects of electric field deformation due to charge accumulation. 
    
    Despite the close relation between the aging and the deposits formation on the electrodes surfaces \cite{vavra_physics_2003, yasuda_new_2003, sauli_gas_2024}, most studies on aging of GEMs have focused on relating performance to functioning time,  whereas just a few experimental works have focused on the formation of deposits \cite{vavra_physics_2003}. Electron microscopy \cite{advanced1} was employed to demonstrate that after use, carbon, oxygen and silicon were detected in the vicinity of the GEM holes, with signal intensity decreasing with the distance from the hole. Carbon and oxygen had their origin speculated by the authors as being products from the gas ionization (in this case study it was a mixture of argon and carbon dioxide), while silicon was traced as contaminant originated in the grease used in rubber seal. 
    
    Here, we employed time-of-flight secondary ion mass spectrometry (ToF-SIMS) to enable new insights on the deposits formation on the surface of GEM electrodes. We report on the analysis of the molecular content on the surface of an damaged GEM and provide evidence that points to organic deposits formation with significant contribution of degradation products of polyimide. Our results demonstrates that the superior chemical specificity enabled distinguishing environmental contamination to the formed deposits, and more importantly, it enables traceability of the deposits to the polyimide degradation process. With ToF-SIMS data, we also show that chromium, used to promote adhesion of copper onto the polyimide, undergoes a diffusion process, resulting in complete obliteration of the layered structure. 
    
    Even though this is a single sample study, performed in a GEM sample with low history records, we believe our results serves as strong evidence for the necessity of reinterpretation of aging studies in light of deposits formation analysis with higher chemical specificity techniques. 

\section{Methods}

    \subsection{Sample history and case study}

        The sample under investigation was a GEM that was subjected to routine operational conditions in CF$_4$ plus argon mixtures until it suffered damage by a spark. It was an integral component of an imaging study utilizing GEMs and optical readouts \cite{brunbauer_combined_2018, brunbauer_radiation_2018}. The foil features holes of 70 $\mu$m with a pitch of 150 $\mu$m and was produced in the CERN workshop. Pristine foils with the same characteristics were employed as a reference to assess initial surface contaminants related to the production process. Since the foil was used multiple times in different detectors assemblies, no history of usage concerning total dose or amplification was recorded. Because of this, the conclusions in terms of aging and longevity obtained in this study are limited. However, what makes this sample eligible to study is the damage caused by the spark burn leading to a short circuit. The spark \cite{chatterjee_spark_2020} may have left (or been triggered by) some trace residues adsorbed at the surface, and the analysis of surface deposits motivates this study. The selection of this specific foil was also motivated by a proof-of-principle demonstration regarding the efficacy of surface analysis techniques in aging deposits at GEM's electrode surface. 

    \subsection{Surface chemical analysis }

        ToF-SIMS is a surface-sensitive analytical technique that provides detailed chemical and molecular information into the composition and spatial distribution of elements and molecular species on solid sample surfaces with high spatial and mass resolution \cite{SurfaceAnalysis, ToFSIMS}. The fundamental principle involves directing a primary ion beam onto the sample surface, inducing the ejection of secondary ions, including molecular fragments. Ejected ions are extracted towards a detector system that records their travel time over a known distance (flight path). Leveraging the principle that lighter ions travel faster than heavier ones in vacuum, the recorded flight times are then converted into mass-to-charge ratios (\textit{m/z}). The primary ion beam can be focused and raster scanned to provide spatially resolved information with micron to nano-scale resolution. ToF-SIMS can provide information of both organic and inorganic materials, with such versatility that there are applications across diverse fields such as materials science, biology, chemistry, and the semiconductor industry, where precise surface information is paramount.

        Large area ToF-SIMS surface chemical analysis of the GEMs was carried out using a ToF-SIMS IV instrument (IONTOF GmbH). Secondary ion mass spectra were acquired in positive and negative ion polarity mode using a pulsed 25~keV Bi$_3^+$ primary ion beam delivering 0.3-pA. The large area maps were constructed by raster scanning the sample stage and combining patches of datasets where the primary ion beam was raster scanned over areas of 400x400 µm\textsuperscript{2} within each patch of the large area maps. The total ion dose was kept under the static limit of 10\textsuperscript{13}~ions/cm\textsuperscript{2}. A low-energy (20~eV) electron flood gun was employed to neutralise charge build up. 

        ToF-SIMS depth profiling of the degraded GEM was carried out using an OrbiSIMS instrument (Hybrid SIMS, IONTOF GmbH). The Orbitrap analyser was not used for this study. The ToF-SIMS data were acquired in positive ion polarity mode in delayed extraction mode by raster scanning a pulsed 30~keV Bi$_3^+$ primary ion beam (delivering 0.28~pA)  raster scanning 250×250~µm\textsuperscript{2} with a pixel size of 1.9~µm at the center of a 350×350-µm\textsuperscript{2} sputter crater formed using a 20 keV Ar\textsubscript{1460}+ beam delivering 3~nA \cite{kawashima_examination_2016,yamamoto_ioninduced_2013,miyayama_removal_2010}. The analysis was performed in the “non-interlaced” mode with a low-energy (20~eV) electron flood gun employed to neutralise charge build-up. 
        
        A cross section of the degraded GEM sample was prepared using an RMC Products PowerTome. A small piece of the degraded GEM was embedded in Agar100 resin (Agar Scientific Company) before treating with DOW CORNING® Z-6040 Silane (Dow Company) to form a hard block. After 48h-drying in the vacuum oven at 60-70$^o$C for the polymerisation, the block face was trimmed by removing thin layers of up to 800 nm each using a glass knife (Leica Knifemaker, Leica Microsystems Company) and a diamond knife (DiATOME Company) to create a trapezoidal face. sectioning was done until the samples was polished to flat cross-sectional surface in the vicinity of a GEM hole. The flat sectioned surface of the block was directly analysed by ToF-SIMS. Chemical mapping of the GEMs cross-section was done using the same primary ion beam. To minimise smearing effects due to the microtome sectioning, The Ar\textsubscript{1460}+ was used to clean the cross-section of the degraded GEM prior to chemical mapping. 
        
        All images, depth profiles and spectra were processed using SurfaceLab 7.1 (IONTOF GmbH). Mass spectra were calibrated in SurfaceLab using organic secondary ions  for organic assignments and inorganic secondary ions for inorganic assignments \cite{green_tof-sims_2006}.

\section{Results and discussions}


    \subsection{Observation of organic material at the surface of degraded GEM}

\begin{figure*}[htb!]
            \centering
            \includegraphics[width=0.9\textwidth]{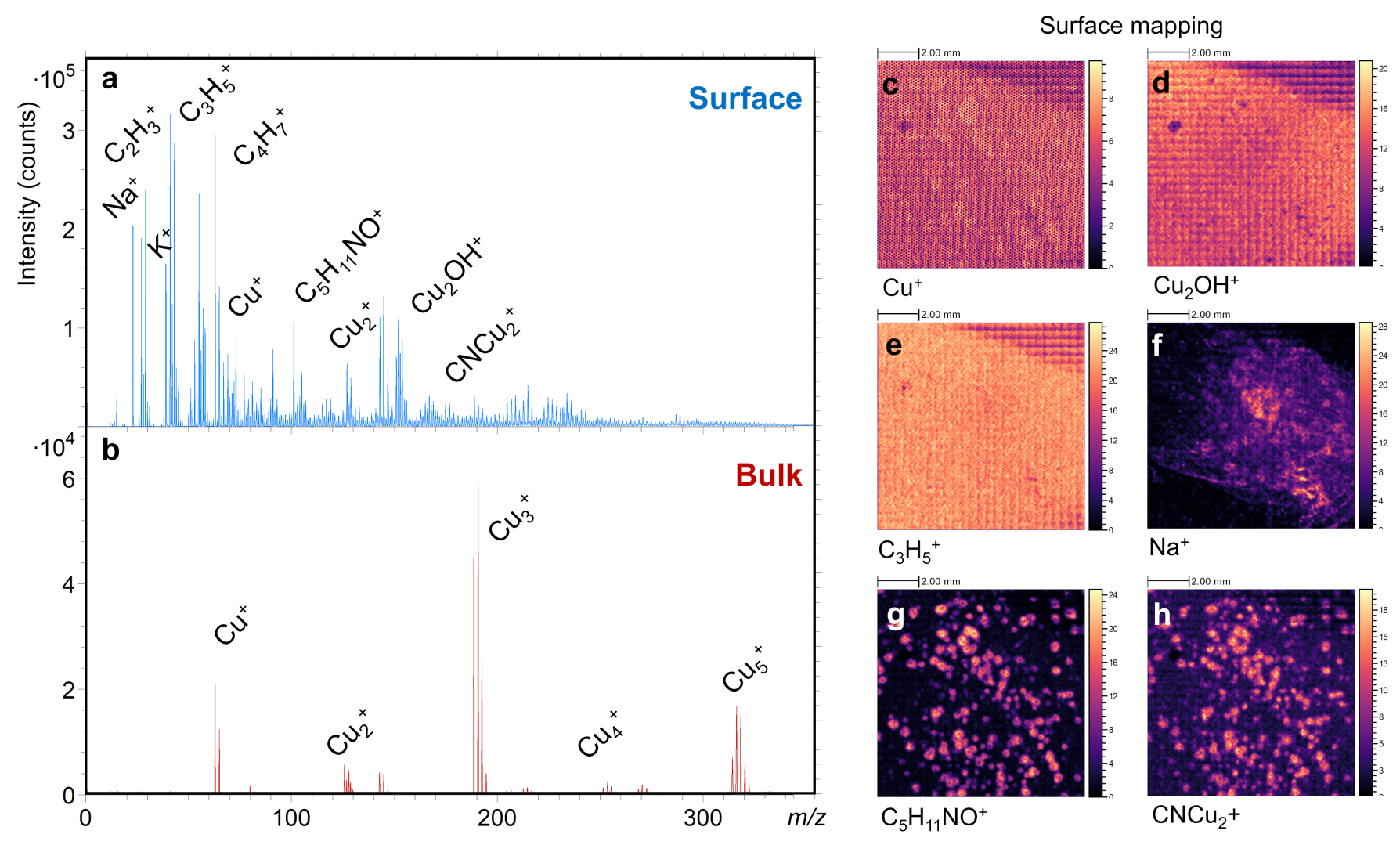}
            \caption{ToF-SIMS surface chemical analysis of degraded GEM. Mass spectra of \textbf{a} surface and \textbf{b} bulk of sample with main secondary ions indicated. The bulk was exposed by etching the surface layer away using a gas cluster ion beam. Chemical maps showing spatial distribution of \textbf{c-h} copper related signal, \textbf{e} environmental contamination (adventitious carbon), \textbf{f} accidental contamination (fingerprint), \textbf{g-h} redeposition of products from avalanche process.}
            \label{fig:patterns}
        \end{figure*}

        In contrast to electron microscopy, ToF-SIMS offers molecular information, which in this particular case, can be more effective in providing information on the origin of the contamination. Moreover, the information is surface sensitive and can detect residues of thickness as thin as a few nanometers, which include surface contamination from different sources (see fig. \ref{fig:patterns}-a,b). The spatially resolved information is important to disentangle environmental contamination (uniformly distributed over the surface, fig. \ref{fig:patterns}-e), to accidental contamination (finger prints, for example, with wide patterns that does not match to the hole patterns, fig. \ref{fig:patterns}-f), and the avalanche-related residues (the signals with spatial distribution that matches the holes distributions and with flight times compatible with masses that we can trace back to materials present in the detector, Fig. \ref{fig:patterns}-g,h). It is also important to note that molecular masses measured by ToF-SIMS are actually fragments of the heavier molecules that are adsorbed onto the surface.

\begin{figure}[htb!]
            \centering
            \includegraphics[width=0.45\textwidth]{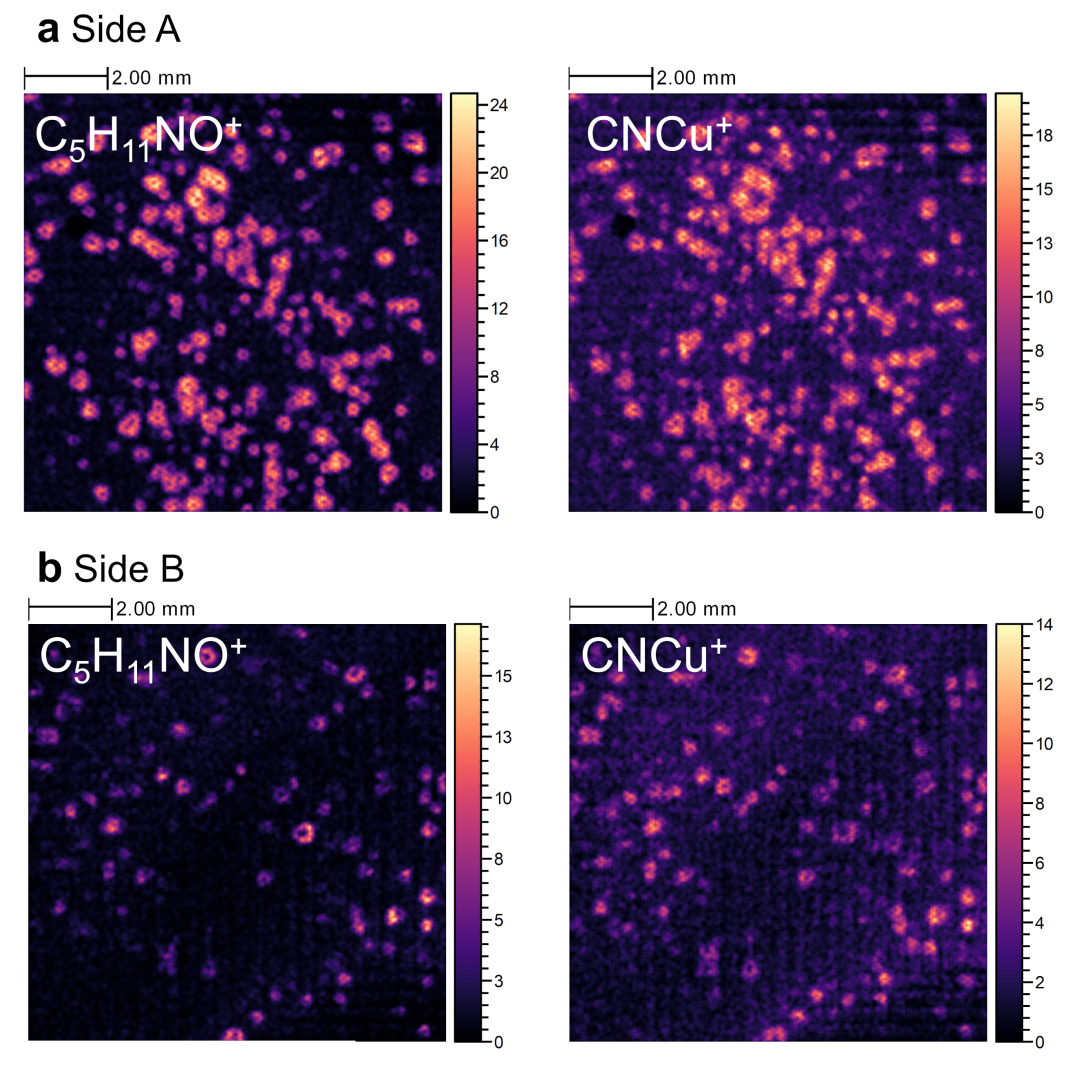}
            \caption{Distribution pattern of C$_5$H$_{11}$NO+ and CNCu$_2^+$ masses on both sides of the GEM foil. The differences may reveal a orientation-dependent redeposition.}
            \label{fig:bothsides}
        \end{figure}

        Two specific secondary ions are important in this case of study, C$_5$H$_{11}$NO$^+$ (\textit{m/z} 101.09 u, $\Delta m$/$m$ accuracy 30.7 ppm) and CNCu$_2^+$ \textit{m/z} 151.85 u, $\Delta m$/$m$ accuracy -95.5 ppm). The presence of nitrogen together with the high molecular mass indicates that their parent molecule likely originate from the polyimide insulating layer (polyimide molecular formula C$_{41}$H$_{22}$N$_{4}$O$_{11}$). The presence of copper atoms in the second molecule shows that the layer is thin enough and indicates an interaction with the copper surface. In this sense, we can hypothesize that these are a result of a redeposition process of polyimide fragments removed from the internal walls of the GEM holes by the avalanche of ionization, offering an alternative explanation to the destination of the material eroded as observed in other studies \cite{alfonsi_studies_2005}. Fig. \ref{fig:bothsides} shows the pattern distribution of both molecules on both sides of the foil. As there is no record of usage of the GEM, conclusions are limited. Nevertheless, differences in the counts gives indication for an orientation-dependent redeposition process.

\begin{figure}[htb!]
            \centering
            \includegraphics[width=0.45\textwidth]{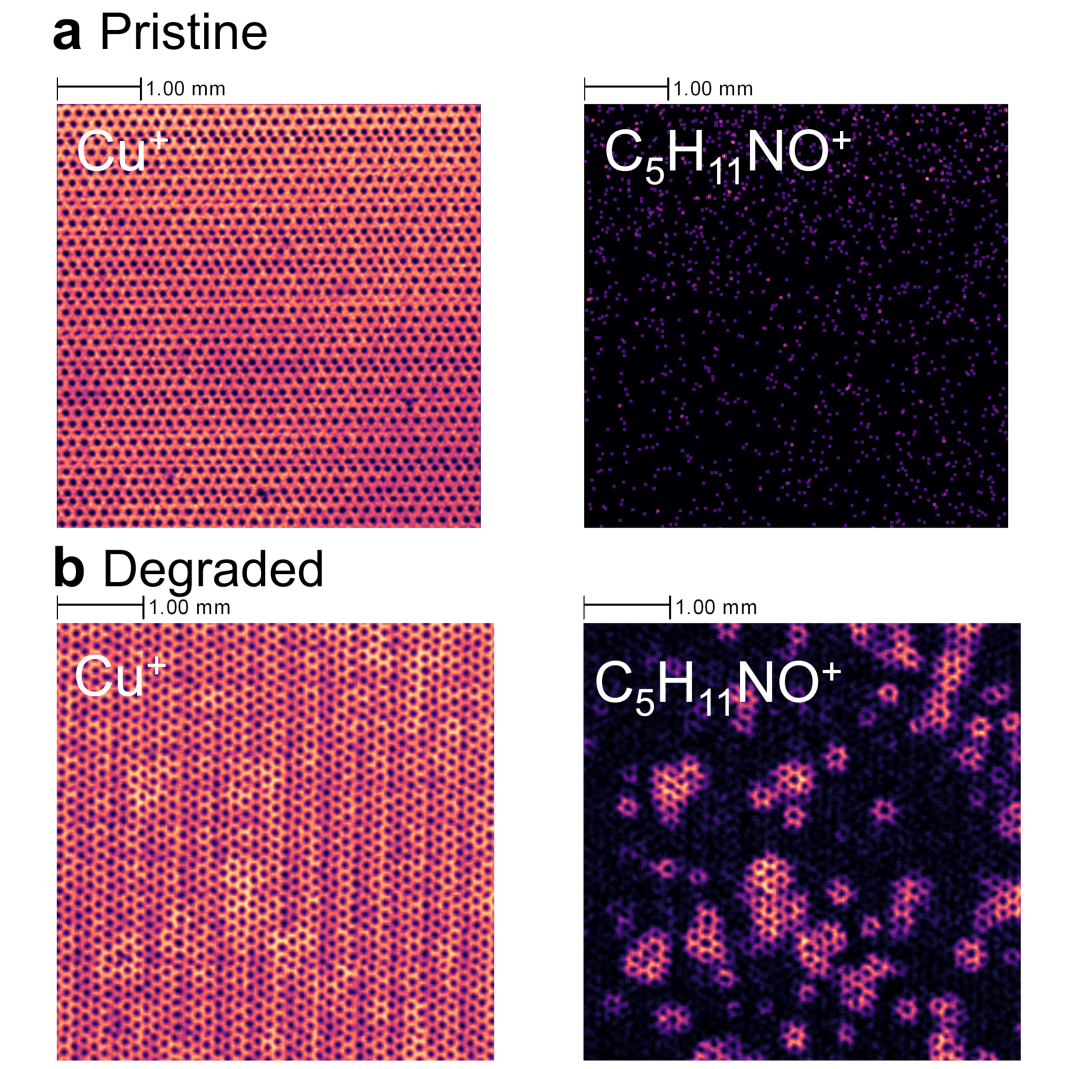}
            \caption{Characteristic organic signal is only detected at the surface of aged GEM. Cu$^+$ and C$_5$H$_{11}$NO$^+$ maps for \textbf{a} pristine and \textbf{b} degraded GEM surfaces. Intensity scales are normalized to the maximum counts within each image. Only statistical noise is observable in the C$_5$H$_{11}$NO$^+$ image for the pristine GEM foil. 
            }
            \label{fig:degradXpristine}
        \end{figure}

        To check if organic content does not originate from the production chain, we also analyzed pristine GEM foils (as provided by CERN workshop). Fig. \ref{fig:degradXpristine} shows that the C$_5$H$_{11}$NO$^+$ signal  and spatial pattern is only detected in the used foil.
        
    \subsection{Source of organic material}
        
\begin{figure}[htb!]
            \centering
            \includegraphics[width=0.45\textwidth]{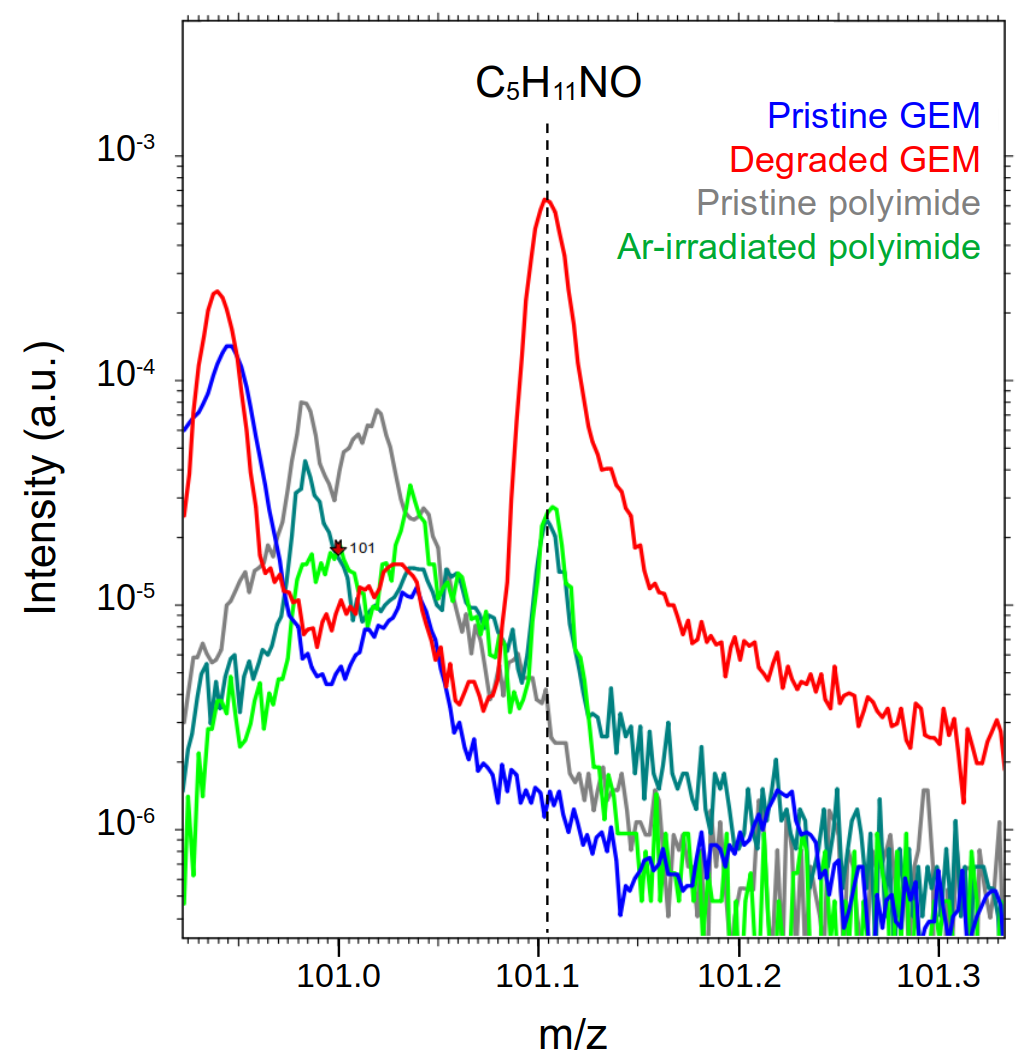}
            \caption{Mass spectra  for pristine and degraded GEM surface, and reference polyimide samples before and after argon irradiation in the region corresponding to C$_5$H$_{11}$NO+ \textit{m/z}. The peak is observed only in the degraded GEM and argon irradiated polyimide sample. This reveals the material redeposition follows a fragmentation of the polyimide molecule.}
            \label{fig:signatures}
        \end{figure}

        To confirm the hypothesis that C$_5$H$_{11}$NO$^+$ originates from polyimide degradation, we analyzed the mass spectra of pristine GEM foils, pristine polyimide foils and argon bombarded polyimide foils (pristine polyimide foils bombarded with 23 keV Ar$^{+}$ ion beam with $10^{15}$ ions/cm$^2$ fluence \cite{trindade_modificacoes_2013}). The comparison of the mass spectra with that obtained with the degraded GEM foil shows that the peak on the mass corresponding to C$_5$H$_{11}$NO$^+$ is not detected in the pristine GEM nor in the pristine polyimide foils. Only the argon irradiated polyimide foil presents the same peak. This comparison can be observed in fig. \ref{fig:signatures}. These results indicate that C$_5$H$_{11}$NO+ is characteristic of a polyimide degradation product, which is induced by Ar-irradiation in the reference polyimide sample. In the case of the GEM foil, possible degradation agents are the radiation incident during the routine use of the detector, or a secondary effect of the electron avalanche process. These should be investigated more carefully in future tests. ToF-SIMS is undertaken under static conditions and using very low beam currents (pA range), so that no degradation is induced during the measurements of pristine samples, as shown in fig. \ref{fig:signatures}.

    \subsection{Fluorine content}

\begin{figure}
            \centering
            \includegraphics[width=0.45\textwidth]{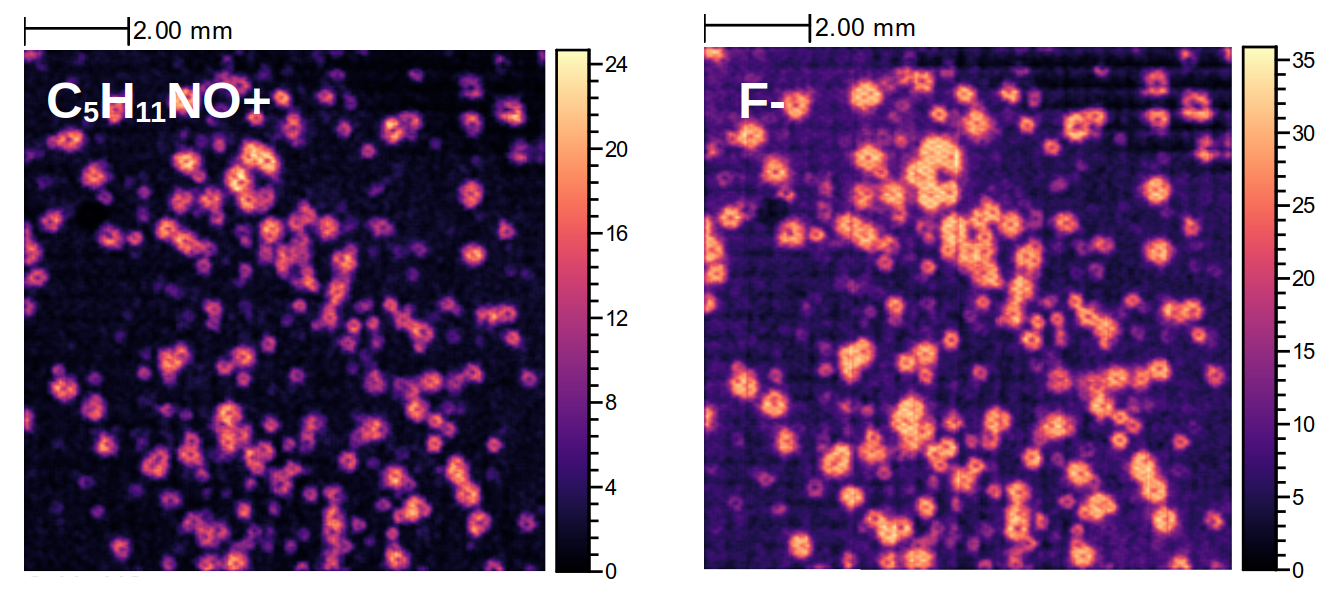}
            \caption{Comparison of C$_5$H$_{11}$NO+ and F- ToF-SIMS maps of the degraded GEM surface.}
            \label{fig:fluorine}
        \end{figure}

        In data obtained with the negative ion collection mode we observed the complete overlap between the $F$$^-$ and C$_5$H$_{11}$NO+ signal distributions. This comparison is shown in fig. \ref{fig:fluorine}. Our result is complementary to that reported in \cite{corbetta_studies_2021} indicating that, not just the fluorine is spread all over the interior of the detector, but its high electron affinity indicates that the fluorine originates in the CF$_4$ breakup as a central player in the observations of this study with supportive evidence that the observed patterns around holes may correlate with avalanche events.

    \subsection{Possible relation to sparks}

\begin{figure*}[htb!]
            \centering
            \includegraphics[width=1.0\textwidth]{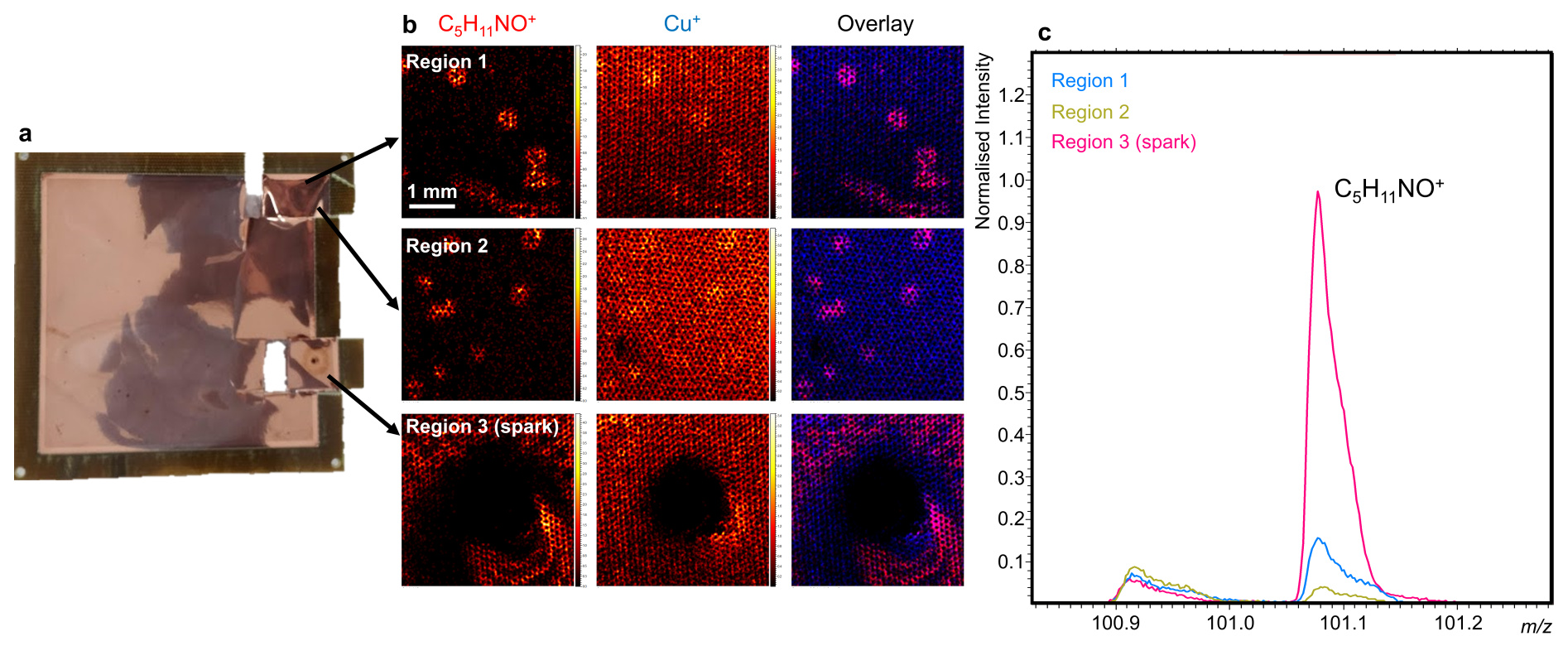}
            \caption{Localization and SIMS chemical analysis of degradation in and aged GEM \textbf{a} optical image of different regions on the degraded GEM foil. \textbf{b} The area containing the spark burn has a larger surface coverage and \textbf{c} exceeds other regions by a factor of 5 to 10. Normalized intensity mass spectra around the C$_5$H$_{11}$NO$^+$. }
            \label{fig:sparks}
        \end{figure*}

        We further investigated the surface chemistry in different regions of the GEM foil, including the region that surrounds the spark burn that damaged the foil (fig. \ref{fig:sparks}-a). We observed a much higher signal intensity for C$_5$H$_{11}$NO+ near the spark (fig. \ref{fig:sparks}-b,c), revealing a possible relation between the spark and the organic deposits on the surface. This relation is yet not clear, but two models seem likely: i) Either the insulating layer promotes electron ejection by the field emission from the dielectric coated metallic surfaces (Malter effect) \cite{zhou_theory_2020}, and as this is thickness dependent, when a critical thickness is reached the electrons are ejected above a threshold to trigger a intense spark, ii) or the spark induced the spread of such contaminant and what we observe in other spots are contamination induced by small sparks not intense enough to damage the foil. Both scenarios are possible and may occur simultaneously, with sparks residues inducing higher intensity sparks, in a chain-process that may lead to GEM failure. 
        
        Note that, the field emission by dielectric coated metallic surfaces is a quantum tunneling process \cite{zhou_theory_2020}. The emission probability of electrons at the Fermi level in the copper presents resonances as the film thickness grows. Thus, when the first resonant thickness is reached, an enhanced emission probability can lead to the stream formation and induction of sparks.
        
\begin{figure*}[htb!]
            \centering
            \includegraphics[width=1\textwidth]{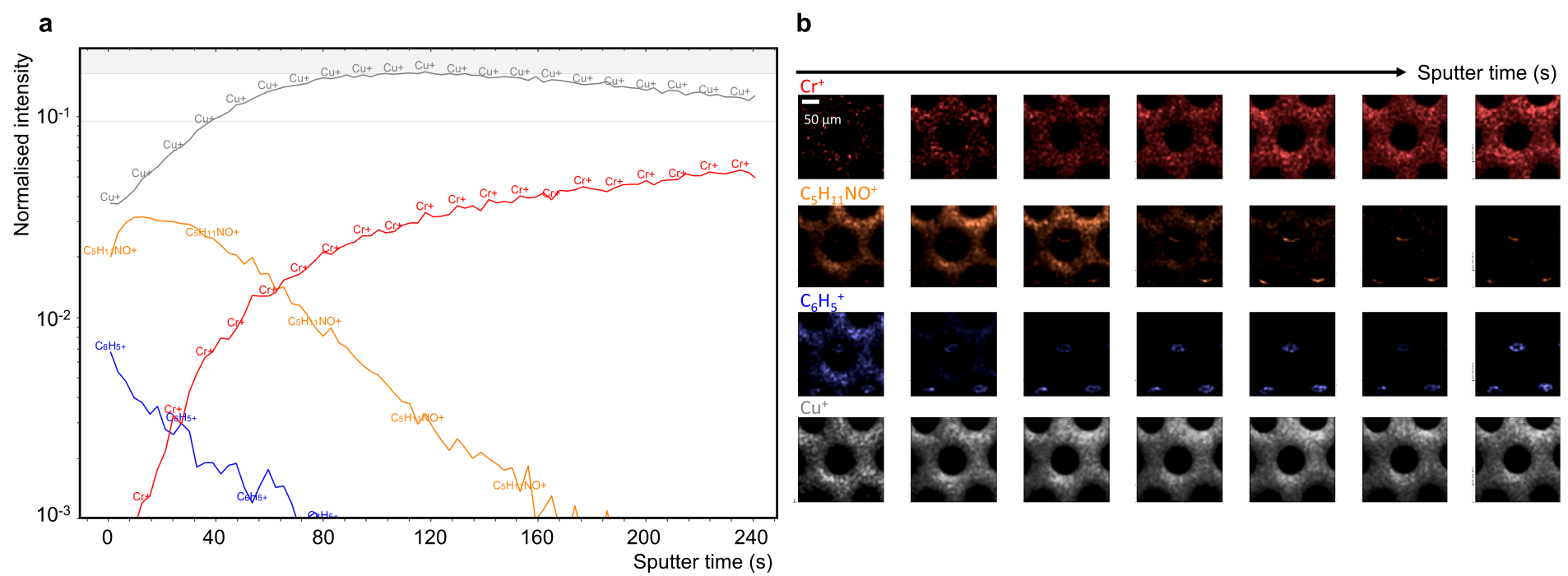}
            \caption{ToF-SIMS depth profiling near the degraded GEM spark region. \textbf{a} Intensity of secondary ions of interest as a function of sputter time (proportional to depth). \textbf{b} Chemical maps of selected scans with sputter time varying from left to right.}
            \label{fig:DP}
        \end{figure*}

        Chemical depth profiling near the spark burn reveals the depth distribution of different signals of organic molecules (fig. \ref{fig:DP}). The signal related to adventitious carbon contamination (C$_6$H$_5^+$)  rapidly reduces, followed by a peak of intensity for C$_5$H$_{11}$NO$^+$. The observation of the same signature as the region not damaged by the spark may be an indication that this was accumulated at the surface in higher levels in the region prior the spark damage. This hypothesis however, requires further investigation. Furthermore, the sub-surface region is dominated by copper-related signals and also shows an increase of chromium which, as detailed in the next section, may have diffused from the interface with polyimide but did not have enough energy to reach the top surface, as typically happens in diffusion phenomena.

    \subsection{Evidence for material migration}

\begin{figure*}[htb!]
            \centering
            \includegraphics[width=0.75\textwidth]{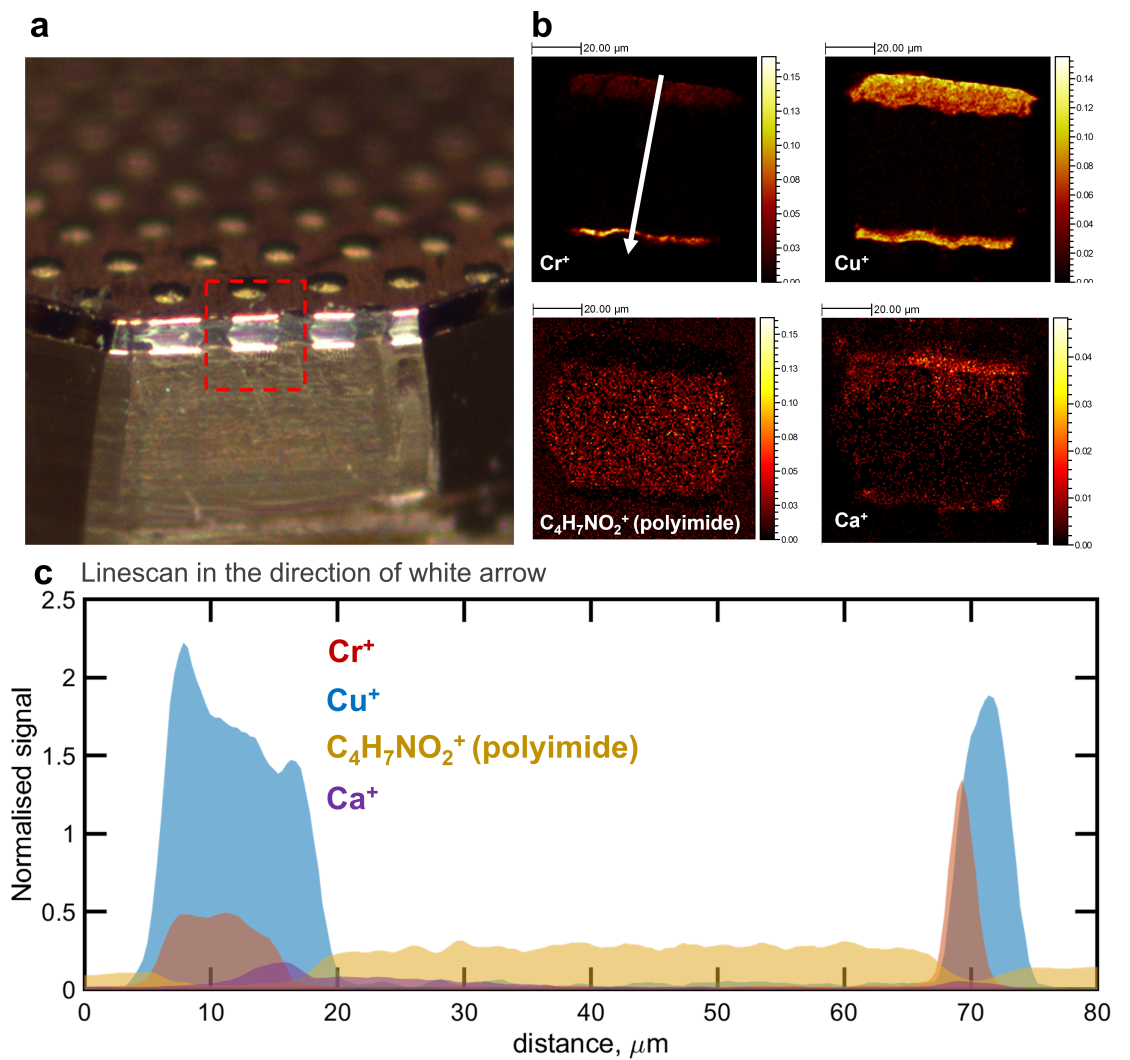}
            \caption{Evidence for material migration. \textbf{a} Photograph of GEM sample embedded in resin with cross-section exposed using a microtome. \textbf{b} ToF-SIMS cross-sectional images showing distribution maps of Cr$^+$, Cu$^+$, C$_4$H$_7$NO$_2^+$ (polyimide signature), and Ca$^+$ signals (clockwise). Chromium is used as an intermediate layer to promote adhesion of copper on polyimide, and calcium is a non-organic filler in the insulating layer. \textbf{c} Intensity profiles along the arrow for the analyzed chemical species. The spread of chromium signal into the copper layer on the right reveals a diffusion process took place during functioning.}
            \label{fig:crmigration}
        \end{figure*}

        Cross-sectional cuts of the degraded GEM foil were produced using a microtome after embedding a piece of the foil into resin.  Fig. \ref{fig:crmigration} shows maps of ToF-SIMS signal distribution for chromium, copper and organic signal characteristic of polyimide (C$_4$H$_7$NO$_2^+$). Chromium is used as an adhesion layer in the interface between the polyimide and the copper \cite{mindur_investigation_2020}. It is possible to observe the chromium signal is spread in the top copper layer, indicating a diffusion process on this side. It is also possible to observe the obliteration of the layered structure in the linescan of fig. \ref{fig:crmigration}. The cross-section surface was cleaned using a gas cluster ion beam (GCIB) to ensure all modifications introduced by the microtome process did not affected the present observations.

        Diffusion is a phenomenon driven by temperature. Thus, by comparison of the chromium signals on both sides of the degraded GEM, it is possible to conclude  that one side reached higher temperatures than the other for a longer time. As polyimide is an excellent thermal insulator, this is plausible. Even though diffusion can also be accelerated by radiation, given the thicknesses involved, it is unlikely that one side received much more radiation than the other, making the pure thermal process hypothesis more believable.

        The data also points to chemical changes in the insulating layer. It is common during the production of polymers that some inorganic fillers are added in the formulation for better cost and performance. In the case of polyimide, calcium-based fillers are regularly adopted. The ToF-SIMS cross sectional maps of Ca+ indicates calcium migration (see panel in fig. \ref{fig:crmigration}-b,c). However, in this case we speculate it is a result of an electric field driven process rather than diffusion. The reason for this difference is that polyimide is a thermal insulator, and that the calcium presented movement even in the colder side (attributed from the chromium migration observation). At least until now we cannot relate this result with the longevity of the GEM foil or its performance in a detector.

\section{Conclusions}
    
    The study of aging and degradation of GEMs is critical for present and future experiments in particle and high-energy physics. The effects include systematic gain and resolution variations, affecting lower threshold limits, calibrations and signal differentiation. The ECFA report \cite{ecfa_detectors_rd_roadmap_process_group_2021_2021} shows that many of the physics goals of future experiments requires improved longevity and higher rate capabilities. In this sense, a deeper understanding of degradation processes occurring in the detector are necessary, with one of these processes being the surface deposit formation during operation. A material science perspective can be helpful in this topic.

    We used ToF-SIMS to probe the surface chemistry of a GEM foil damaged by a spark burn after routine use in argon mixtures with CF$_4$ in studies with optical readout, and observed a signal related to organic deposits on the copper surface. The presence of nitrogen in these molecules establish an origin relation to the polyimide layer of the GEM. The higher signal for the organic molecules in the region of the spark burn is an indication of a connection with sparks, with two main hypotheses: either the organic layer is inducing the ejection of electrons by the Malter effect until it provide conditions to spark formation, or the organic layer is a result of countless micro-size sparks until one large spark caused the permanent damage. This requires further investigation.

    Chemical maps of a cross section indicated chromium migration by diffusion, with complete loss of the layer responsible to promote the adhesion between the copper layer to the polyimide on one side of the GEM foil. This may be an indication of thermal anisotropy, and to additional long term effects. Possible consequence of this is the detachment of the copper layer in scales. This subject is unexplored, and more studies are necessary to determine if this is relevant or not in the functioning of a detector. It is known, for example, that copper films present some reactivity with polyimide, and the application of potential differences can favor interdiffusion, or even the establishment of chemical bonds between copper and polymer constituents (C, O and N) \cite{lee_aging_2013}. It may produce some consequences in much longer functioning times.

    Our results reveals that surface-sensitive chemical mapping methods such as ToF-SIMS can provide important insights on the deposits formation that is related to aging and degradation processes of a GEM. However, more controlled and systematic experiments are needed to correlate our findings to aging and extend the conclusions. History records, together with dose and amplification monitoring are necessary to a more general result. Nevertheless, even with the lack of this information, the present study is important to call the attention for the need of better chemical specificity of organic molecules when studing the surface of GEMs, which may provide key information to improving performance and lifetime of GEMs.

\section*{Acknowledgments}

    This research was supported by the funding agencies CNPq and FAPESP (grant numbers 406672/2022-9, 306414/2022-8, 2019/07426-0, and 2022/03043-1), and EPSRC (grant numbers EP/P031684/1 and EP/P029868/1). The authors also acknowledge access to facilities at the Nanoscale and Microscale Research Centre (nmRC) of the University of Nottingham. 





\bibliographystyle{model1-num-names}
\bibliography{references.bib}







\end{document}